\documentclass[runningheads,fleqn]{svmult}
\usepackage{makeidx}   % allows index generation
\usepackage{graphicx}  % standard LaTeX graphics tool
\usepackage{subeqnar}  % subnumbers individual equations
\usepackage{multicol}  % used for the two-column index
\usepackage{taphys}    % flushleft layout of captions etc.
\makeindex             % used for the subject index
\def\tauphi{$\tau_{\phi}$ }

\begin{document}
\title*{Electron Coherence in Mesoscopic Kondo Wires}
\toctitle{Electron Coherence in Mesoscopic Kondo Wires}
% allows explicit linebreak for the table of content
%
%
\titlerunning{Electron Coherence in Mesoscopic Kondo wires}
% allows abbreviation of title, if the full title is too long
% to fit in the running head
%
% \author{F\'elicien  Schopfer\inst{1}
% \and Christopher B\"auerle \inst{1}
% \and Wilfried Rabaud\inst{1}
% \and Laurent Saminadayar\inst{1}}
% \author{F\'elicien Schopfer, Christopher B\"auerle$^{*}$, Wilfried
% Rabaud$^{\dagger}$ and Laurent Saminadayar$^{\ddagger}$}
\author{F. Schopfer, C. B\"auerle$^{*}$, W. Rabaud$^{\dagger}$ and L.
Saminadayar$^{\ddagger}$}

\authorrunning{F\'elicien Schopfer \textit{et al.}}
% if there are more than two authors,
% please abbreviate author list for running head
%
%
\institute{Low Temperature Research Laboratory, CRTBT-CNRS, 
B.P. 166 X, 38042 Grenoble Cedex 09, France}

\maketitle              % typesets the title of the contribution

\begin{abstract}

We present measurements of the magnetoresistance of long and narrow
quasi one-dimensional gold wires containing magnetic iron impurities. 
The electron phase coherence time extracted from the weak antilocalisation
shows a pronounced plateau in a temperature region of
$300\,$mK\,-\,$800\,$mK, associated with the phase breaking due to the
Kondo effect.  Below the Kondo temperature, the phase coherence time
increases, as expected in the framework of Kondo physics.  At much
lower temperatures, the phase coherence time saturates again, in
contradiction with standard Fermi liquid theory.  In the same
temperature regime, the resistivity curve displays a characteristic
maximum at zero magnetic field, associated with the formation of a
spin glass state.  We argue that the interactions between the magnetic
moments are responsible for the low temperature saturation of the
phase coherence time.

\end{abstract}

\section{Introduction}

The understanding of the ground state of an electron gas at zero
temperature is one of the major challenges in Solid State Physics. 
For a long time it has been known that such a ground state is well described
by Landau's theory of Fermi liquids \cite{pines+nozieres}.  In this
description, the lifetime of quasiparticles is infinite at zero
temperature, as the coupling to the environment tends to zero.

Alternatively, in mesoscopic physics, one key physical concept is the
phase coherence time, \textit{i.e.} the time an electron can travel in
a solid before it looses its phase coherence and thus its quantum,
wave like behaviour.  Such a decoherence is due to inelastic
processes, like electron-phonon, electron-electron or electron-photon
collisions.  It has been shown by Altshuler and
coworkers \cite{altshuler} that the phase coherence time diverges at
zero temperature as electron-phonon, electron-electron and
electron-photon interactions all go to zero at zero temperature.

However, recent experiments on metallic as well as semiconductor wires
suggest that the phase coherence time saturates at very low
temperature \cite{mohanty_prl_97}.  Following this work, it has been
argued that the observed saturation is indeed universal and intrinsic,
and due to electron-electron interactions in the ground state of the
Fermi liquid \cite{zaikin}. Contrary to this, other interpretations
argue that this saturation is extrinsic and due to the coupling to
other degrees of freedom, like two level systems \cite{tls}.  On the
other hand, some experimental results suggest that the dephasing
depends on the dimensions of the samples \cite{natelson_prl_00},
whereas another group argues that some of their experimental results 
agree with standard theory \cite{saclay}, at least down to 50\,mK. 
It should be noted,
however, that the problem of relaxation at zero temperature is not
subject of debate in the mesoscopic community only.  Recent
experiments in spin polarized helium, a textbook example of a Fermi
liquid, show a saturation of the transverse spin-diffusion coefficient
at zero temperature \cite{akimoto}.  This equally raises key questions
about the applicability of conventional Fermi liquid theory.

Recent experiments invoke the coupling to magnetic impurities as a
possible source of the frequently observed low temperature saturation
of the phase coherence time \cite{birge,schopfer,anthore}.  It is well
known that in metals the interaction of conduction electrons with
magnetic impurities gives rise to the Kondo effect \cite{kondo}. 
Concerning transport properties of metals, the best known feature
of this effect is the existence of a minimum and a subsequent
logarithmic increase of the resistivity with decreasing temperature
below the Kondo temperature $T_{K}$.  The influence of Kondo
impurities on the dephasing rate, on the other hand, is by far more
subtle.

Finally, it is well known that above a certain amount of impurities,
and below a certain temperature, RKKY interactions between magnetic
moments lead to the formation of a spin glass \cite{mydosh}. 
This regime  has basically not been explored so
far and may contain a great deal of new physical phenomena.

All this physics related to magnetic impurities leads to new energy
scales: the Kondo temperature $T_{K}$ and the spin
glass transition temperature $T_{g}$.  Both energy scales have to be considered
when dealing with the ``zero'' temperature limit, and have also to be
introduced in the theoretical description of dephasing in
mesoscopic wires.

\section{Historic}

Already in the early days of weak localisation, many experimentalists
observed a systematic saturation of the electron phase coherence at
low temperatures, when extracted from low field
magnetoresistance \cite{gershenzon,rosenbaum}.  This saturation has
often been attributed to the presence of some residual magnetic
impurities \cite{bergmann_84}.

To our knowledge, the first measurements which clearly demonstrated
the strong influence of magnetic impurities on the phase coherence,
even in the presence of extremely dilute magnetic impurities (below
the ppm level) has been carried out by Pannetier and coworkers in the
80$^{ths}$ \cite{pannetier_prl,pannetier_physica}.  These measurements
have been performed on extremely pure Au samples and coherence lengths
of several micrometers have been obtained at low temperatures.  Again
in these experiments, the phase coherence time was almost temperature
independent below 1 Kelvin.  By annealing the samples, the authors
could show that the phase coherence time increases substantially.  The
annealing process oxidizes magnetic impurities and hence suppresses
decoherence due to the Kondo effect.  These experiments therefore
clearly show that the presence of an extremely small amount of
magnetic impurities can lead to substantial electron decoherence at
low temperatures.

A different method to suppress the effect of magnetic impurities can
be achieved by applying a sufficiently high magnetic field in order to 
fully polarise the magnetic impurity spins.  In this case, weak 
localisation measurements are not possible to extract the phase coherence time. 
On the other hand, measurements of Aharonov Bohm (AB) oscillations and universal
conduction fluctuations (UCF) are possible.  Pioneering work on both,
UCF and AB oscillations in quasi 1D quantum conductors containing a
small amount of magnetic impurities (down to 40ppm) has been performed
by Beno\^\i t and coworkers in the late 80$^{ths}$ \cite{benoit}.  In 
this work the authors could clearly show that UCF as well as AB
oscillations increase considerably at fields larger than 1 Tesla,
showing the suppression of the Kondo effect due to the polarization of
the magnetic impurity spins.  In the context of the present debate on
the low temperature saturation of $\tau_{\phi}$, these measurements have
been repeated recently on metallic samples containing more
dilute magnetic impurities \cite{birge}.

In this article we point out another effect which leads to a
saturation of \tauphi at low temperatures when measured by weak
localisation, namely the formation of a spin glass state.  We show
that, even in the presence of very dilute magnetic impurities, the
impurities cannot be regarded as independent (single impurity limit)
at low temperatures and interactions between the magnetic impurities
have to be taken into account.  It is well known that RKKY
interactions between magnetic impurities lead to the formation of a
frozen spin configuration at a characteristic temperature $T_g$.  In
systems containing very dilute magnetic impurities, this temperature
lies well below the Kondo temperature, hence sets another energy
scale, which can be of the order of the lowest temperatures presently
accessible in experiments.

\section{Experimental}

In this article we report on measurements of the temperature
dependence of the low field magnetoresistance and resistivity of
quasi one-dimensional (1D) long and narrow Au/Fe Kondo wires down to
temperatures below 0.1 $T_K$.

Sample fabrication is done using electron beam lithography on silicon
substrate.  The metal is deposited with a Joule evaporator and
standard lift-off technique.  In order to improve adhesion to the
substrate a $1\,$nm thin titanium layer is evaporated prior to the
gold evaporation.  Two sources of $99.99\%$ purity with different iron
impurity concentrations are employed for the gold evaporation.  The
actual iron impurity concentration is determined \textit{via} the
resistance variation at low temperature due to the Kondo effect.  Such
a method directly characterises the purity of the samples, which may
be quite different from the purity of the sources.

The samples (A and B) have the same geometrical parameters: their
lengths, widths and thicknesses are $L=450\mu m$, $w=150nm$ and
$t=45nm$.  The 1 K resistance value for sample A (B) is 4654$\Omega$
(2235$\Omega$).  The samples are quasi 1D with respect to both, the
phase-breaking length $l_\phi$\,=\,$\sqrt{D\tau_{\phi}}$ and the
thermal length $l_T$\,=\,$\sqrt{\hbar D/k_{B}T}$, $D$ being the
diffusion constant.  From the relation $D$\,=\,1/3\,$v_{F}\,l_{e}$, we
obtain a diffusion coefficient of $5.6\,10^{-3}\,m^2/s$ and $11.5\,10^{-3}\,m^2/s$ for
sample A and B, respectively.

\begin{figure}[h]
\centerline{\includegraphics*[width=7.0cm]{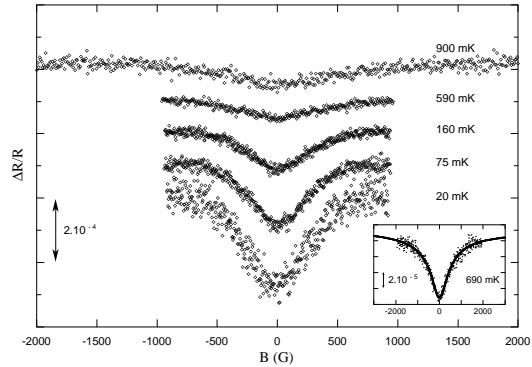}}
\caption{Low field magnetoresistance of sample A at different 
temperatures. Inset shows a fit of weak localisation theory to the 
data}
\label{magneto_AA}
\end{figure}

\begin{figure}[h]
\centerline{\includegraphics*[width=7.0cm]{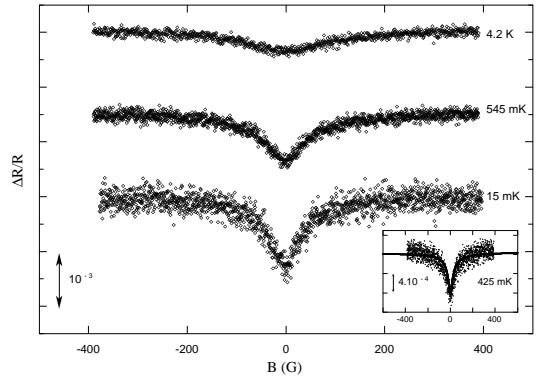}}
\caption{Low field magnetoresistance of sample B at different 
temperatures. Inset shows a fit of weak localisation theory to the 
data}
\label{magneto_AB}
\end{figure}

\begin{figure}[h]
\centerline{\includegraphics*[width=6.5cm]{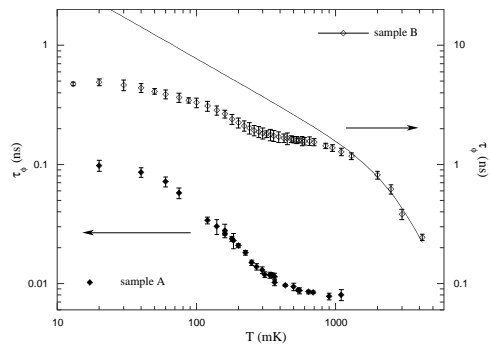}}
\caption{Phase coherence time as a function of temperature. The solid 
line is the theoretical prediction from \cite{altshuler} for 
sample B}
\label{tau_phi}
\end{figure}

We have studied very carefully the temperature dependence of the low
field magnetoresistance as well as the electrical resistivity over a
temperature range extending from 4.2\,K down to 15\,mK. Figures
\ref{magneto_AA} and \ref{magneto_AB} display the magnetoresistance at
different temperatures for two samples containing different amounts of
magnetic iron impurities.  From fits to weak localisation theory for
quasi 1D conductors (see insets) we extract the phase coherence length
$l_{\phi}$.  For the fitting procedure, we first determine the spin-orbit 
scattering length at temperatures above 0.5\,K with
magnetoresistance curves covering a field span of $\pm$\,2000 G. We
then fix the spin-orbit scattering length to the obtained value of
50\,nm for all weak localisation fits \cite{comment1}.  Using the
measured geometrical and electrical parameters of the samples, the
only fitting parameter is hence the phase coherence length $l_{\phi}$. 
From the relation \tauphi = $l_{\phi}^{2}/D$, we then obtain the
phase coherence time \tauphi as shown in Fig. \ref{tau_phi}.

Three distinct temperature regimes are clearly distinguishable.  At
high temperatures (above 1K), the phase coherence time decreases
rapidly with increasing temperature due to electron-phonon coupling. 
This temperature dependence is well described by a $T^{-3}$ power law,
as expected from theory.  At temperatures between 0.3\,K and 1\,K the
phase coherence time shows a pronounced plateau.  Here the temperature
independence of \tauphi is caused by dephasing due to the Kondo effect
as we will see later, when we discuss the temperature variation of the
resistivity.  Below 0.3\,K, the phase coherence time increases again,
because of the partial screening of the magnetic
impurities \cite{mohanty_kondo}.  At lower temperatures, however, we
again observe an apparent saturation of $\tau_{\phi}$.  In order to
understand this rather unusual temperature dependence of
$\tau_{\phi}$, it is important to analyse the temperature dependence
of the electrical resistivity.

\begin{figure}[h]
\centerline{\includegraphics*[width=6.5cm]{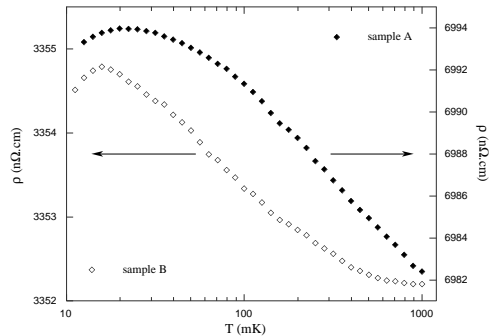}}
\caption{Resistivity variation as a function of temperature, 
measured at zero magnetic field}
\label{R_T_AA}
\end{figure}

Figure \ref{R_T_AA} displays the temperature variation of the
resistivity for both samples in zero magnetic field.  The total
contribution to the resistivity is given by three different
contributions: electron-electron interaction, weak localisation and
the contribution due to magnetic impurities.  
Below 1\,K one observes an increase of the
resistivity for both samples.  This increase is due to the Kondo
contribution and electron-electron interaction.  At the lowest
temperatures, one observes a clear maximum in the resistivity for both
samples.  Since the amplitude of the weak localisation curves is
basically temperature independent at these temperatures and since the
electron-electron contribution increases monotonically with decreasing
temperature $\sim$ (1/$\sqrt{T}$), the maximum in the resistivity has
to be due to some other phenomenon related to the presence of magnetic
impurities.

The maximum in the resistivity is a common feature for Kondo systems
\cite{larsen}.  At low temperature, magnetic impurities interact
\textit{via} the RKKY interaction.  In systems with high Kondo
temperatures, and very low impurity concentration, a complete Kondo
screening of the magnetic impurities can be obtained.  This case is
often referred to as the unitary limit, where RKKY interactions are
suppressed.  However, if the concentration is high enough, and the
Kondo temperature low enough, the screening length may be very large
as it varies like $1/T_{K}$.  RKKY interactions are then important and
the magnetic impurities cannot be treated in the single impurity
limit.  The unitary limit is hence never reached, and the system
transits into a spin glass state at a temperature $T_{g}$.  The most
common features of this transition is a maximum in the resistivity
curve as well as an anomaly in the magnetic susceptibility which
appear roughly at $T_{g}$ \cite{comment_max}.  Both phenomena have been
extensively studied in the past: the dependence of the temperature of
the resistivity maximum \cite{laborde_ssc_71,venkat_sg,venkat_proxi_sg}
and of the susceptibility anomaly \cite{frossati} as a function of the
impurity concentration in Au/Fe systems as well as many others. 
However, the effect of such a peculiar spin configuration on the phase
coherence time has not been explored so far \cite{vegvar}.  To our
knowledge, this is the first time that weak localisation measurements
are accessible in this spin glass regime.

In order to determine the impurity concentration of our samples, we
subtract the measured contribution due to weak localisation and fit
the temperature variation of the resistivity at zero magnetic field 
to the following
expression:
\begin{equation}
\rho(T) - \rho_{0}= \frac{\alpha}{\sqrt{T}} 
+\beta\{0.743+0.332\,[1-\frac{ln(T/T_K)}{ln^2{T/T_K} + \pi^2S(S+1)}]\}
\end{equation}
where the first term corresponds to the electron-electron contribution
and the second term to the Hamann expression \cite{hamann} for the
Kondo contribution, with $\beta$ being the impurity concentration in
ppm, $S$ the impurity spin and $\rho_{0}$ the residual resistivity at 
1\,K.  
Taking S\,=\,3/2, T$_{K}$\,=\,300\,mK \cite{laborde_ssc_71} and 
fitting both data sets over the
same temperature range, we obtain an impurity concentration of
approximately 60 ppm (15 ppm) and a coefficient 
$\alpha$\,=\,1.4\,n$\Omega$$\cdot$cm$\cdot$mK$^{-1/2}$ 
(9.5\,n$\Omega$$\cdot$cm$\cdot$mK$^{-1/2}$) 
for sample A (B), compared to the
theoretically expected value of 
36.4\,n$\Omega$$\cdot$cm$\cdot$mK$^{-1/2}$ 
(12.1\,n$\Omega$$\cdot$cm$\cdot$mK$^{-1/2}$). 
The poor agreement between experimental and theoretical
values for coefficient $\alpha$ for sample A is due to our
choice of fitting both sets of data over exactly the same temperature
range.  If we fit the data of sample A over a limited temperature
range ($>$ 100mK), we then recover the theoretically expected value 
for $\alpha$. 
This proves again that in sample A, where the impurity concentration
is higher than in sample B, RKKY interactions between magnetic
impurities are already present at these 
temperatures. As a consequence, the resistivity deviates strongly from 
the Kondo model \cite{comment_alpha}.

The saturation of \tauphi and the subsequent desaturation at lower
temperatures can be well understood in terms of the Kondo effect. 
Spin flip scattering due to the presence of magnetic impurities causes
very efficient dephasing at temperatures around $T_K$.  At lower
temperatures the magnetic impurities become screened by the
surrounding conduction electrons and the spin flip scattering process
is attenuated.  As a consequence, \tauphi increases with decreasing
temperatures \cite{mohanty_kondo}.  At low enough temperatures 
standard Fermi liquid
theory \cite{nozieres} should again describe the temperature dependence
of $\tau_{\phi}$.  It should therefore follow a power law
$T^{-2/3}$ \cite{altshuler} as shown by the solid line in Fig.
\ref{tau_phi}.  This is clearly not the case for our experimental
data.

What can be the origin of the observed low temperature saturation of
$\tau_{\phi}$?  One explanation, however presently very
controversial, is the possible existence of zero temperature dephasing
due to electron-electron interactions \cite{zaikin}.  The agreement of
the experimental data with this theory is reasonable.  For a detailed
comparison of our data with this theory, we refer the reader to
ref. \cite{schopfer}.

Another possibility for the low temperature saturation of \tauphi is
the presence of another type of magnetic impurity with a Kondo
temperature below the measuring temperature (\textit{e.\,g.} Mn).  In 
this case
one would expect again a plateau for \tauphi at temperatures around
$T^{\prime}_{K}$, in qualitative agreement with the data on the phase 
coherence. 
If so, this additional Kondo contribution should also lead to an
increase of the resistivity at low temperatures.
 
None of these two possibilities does explain the maximum
in the resistance curve.  As already mentioned above, the maximum in
the resistance curve is a well known feature which is attributed to
freezing of the magnetic impurities into a spin glass
state. It is thus clear that RKKY
interactions between magnetic impurities are important in our samples
and have to be taken into account in the interpretation of our 
experimental data.

\begin{figure}[h]
\centerline{\includegraphics*[width=7.0cm]{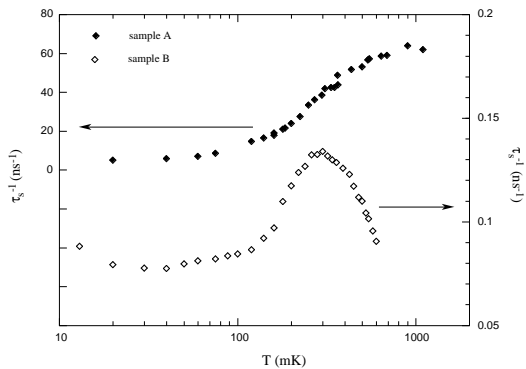}}
\caption{Magnetic scattering rate for sample A and B obtained by
subtraction of the standard dephasing rate from the data of fig. \ref{tau_phi}}
\label{1_tau_ss}
\end{figure}

For this purpose, we extract the spin scattering rate from the
measurement of the phase coherence time.  The total dephasing time is
given by \cite{haesendonck_prl_87}
\begin{center}
\begin{equation}
\frac{1}{\tau_{\phi}}=\frac{1}{\tau_{nm}}+\frac{2}{\tau_{s}}
\end{equation}
\end{center}
where $\tau_{s}$ is the spin scattering rate and $\tau_{nm}$ is the
non-magnetic scattering rate given by the usual formula
\begin{center}
\begin{equation}
\frac{1}{\tau_{nm}}=AT^{2/3}+BT^{3}
\end{equation}
\end{center}
Coefficient A\,=\,0.8\,(0.6)\,ns$^{-1}$\,K$^{-2/3}$ is calculated
using the parameters of sample A (B) and coefficient
B\,=\,0.04\,(0.04)\,ns$^{-1}$\,K$^{-3}$ is obtained by fitting the
data at high temperature. The fit for sample B is diplayed in Fig.
\ref{tau_phi}.  This non-magnetic part of the dephasing time is then
subtracted from our data, and we obtain the spin scattering time as a
function of temperature.  This is displayed in Fig. \ref{1_tau_ss}. 

Both curves exhibit a clear maximum around $T_{K}$ 
\cite{haesendonck_prl_87,bergmann_prl_87}, where the 
dephasing mechanism due to the Kondo effect is the most efficient.
This is associated with the plateau observed around $T_{K}$ in the 
$\tau_{\phi}(T)$ curve \cite{comment_tauphi}. 
Below $T_{K}$, magnetic impurities get screened, and one
expects a decrease of the dephasing time.  This is indeed what is
observed.  Theoretical predictions lead to a $T^2$ behaviour in
Nozi\`eres's Fermi liquid theory \cite{nozieres}, whereas recent work
leads to a 1/ln$^2(T_K$/$T$) dependence for partially screened
impurities \cite{glazman}.  None of these predictions is observed.  The
key point is that for both samples the spin-scattering rate saturates
and is basically constant down to the lowest temperatures.  This is not
surprising: it is well known that spin-spin correlations have a strong
influence on the measured dephasing time. Freezing of magnetic moments
into a spin glass state violates the time reversal symmetry, hence 
leading
to a very efficient dephasing mechanism.  When comparing with the
resistivity curves, it is obvious that this new regime appears around
$T_{g}$.  This saturation in the spin scattering rate can thus be
associated with the formation of a spin-glass due to the RKKY
interactions between magnetic impurities.

\section{Conclusions}

Our results clearly show that RKKY interactions, associated with the
spin glass freezing, lead to a constant spin scattering rate, and 
hence yield a finite phase coherence time at very low temperatures.
It is thus important to consider both energy scales, $T_{K}$ as well as 
$T_{g}$ when dealing with metallic systems containing 
even a very small amount of magnetic impurities.
The understanding of electron dephasing in the temperature range below
$T_{K}$ is certainly a challenge for theory, but is probably the key
point to interpret properly the experiments carried out on metals as
they often contain magnetic impurities on the ppm level at best, with
Kondo temperatures in the mK range.  Measurements at lower
temperatures on samples with very low concentrations of magnetic impurities,
well in the unitary limit, would also be of great interest.  In this
case all magnetic impurities are completely screened, and the standard
Fermi liquid behaviour should be recovered.  This would be the key
test to discriminate between intrinsic dephasing and dephasing due to
Kondo impurities.

\section{Acknowledgements}

We gratefully acknowledge L.I. Glazman, M.G. Vavilov, A. Zaikin,
L.P. L\'evy, O. Laborde, J. Souletie, P. Mohanty, H. Pothier,
A. Beno\^\i t and F. Hekking for fruitful discussions.  
Samples have been made at NanoFab, \textsc{CRTBT-CNRS}.
Part of this work has been performed at the
\textit{Ultra-Low Temperature Facility-University of Bayreuth}
within a TMR-project of the European Community (ERBFMGECT-950072). We are
indebted to G. Eska, R. K\"onig and I. Usherov-Marshak for their
assistance.
\\
\textit{\bf Note added:}
A recent paper by Vavilov et al.\cite{vavilov} predicts a saturation
of \tauphi in the presence of RKKY interactions, in 
agreement with our measurements. 
The disappearance of the resistivity  maximum for 
low impurity  concentrations ($T_{sg} \ll T_K$) as predicted by this
theory,  however, seems to be in contradiction with the experimental data and
calls for a systematic study of the resistivity maximum as a function
of the impurity concentration.

%INDEX%%%%%%%%%%%%%%%%%%%%%%%%%%%%%%%%%%%%%%%%%%%%%%%%%%%%%%%%%%%%%%%
% Please code your entries to include a "mutual" subject index in the
% standard syntax. For your own purposes you may print your
% "personal" index by using the following commands:
%
%\clearpage
%\addcontentsline{toc}{section}{Index}
%\flushbottom
%\printindex
%%%%%%%%%%%%%%%%%%%%%%%%%%%%%%%%%%%%%%%%%%%%%%%%%%%%%%%%%%%%%%%%%%%%%


\begin{thebibliography}{8.}
\bibitem{comment1}This is somewhat oversimplified, since the
spin-orbit scattering length may vary slightly with temperature.
\bibitem{comment_max}The maximum in the resistance curve is a 
precursor of the spin glass transition. The actual transition is situated at a 
slightly lower temperature than this maximum.
\bibitem{comment_alpha}If we subtract the theoretically expected
contribution for e-e interactions from the total resistivity, one
obtains a maximum in the resistivity at a temperature around 60 mK.
\bibitem{comment_tauphi}The saturation value of \tauphi around the 
Kondo
temperature differs very much for the two samples.  This is somewhat
surprising since the impurity concentration is only by a factor of 4
different when extracted from the resistivity.  
The determination of
the impurity concentration via the phase coherence time at around
$T_{K}$ \cite{haesendonck_prl_87} would lead to an impurity
concentration of less than 1ppm for sample B. This is clearly
incompatible with the observed resistance variation as well as the 
position of the maximum in the R(T) curve, which is in relatively good agreement    
with data on the bulk Au/Fe Kondo system (see ref. \cite{laborde_ssc_71}).
This discrepancy is presently not understood.
\end{thebibliography}
\end{document}